\documentclass[a4paper,fleqn]{cas-dc}

\usepackage[numbers,sort&compress]{natbib}

\def\tsc#1{\csdef{#1}{\textsc{\lowercase{#1}}\xspace}}
\tsc{WGM}
\tsc{QE}
\tsc{EP}
\tsc{PMS}
\tsc{BEC}
\tsc{DE}

\usepackage{subcaption}
\usepackage{float}
\usepackage[section]{placeins}
\usepackage{units}
\usepackage{stfloats}

\begin{document}
\let\WriteBookmarks\relax
\def\floatpagepagefraction{1}
\def\textpagefraction{.001}
\shorttitle{A Laser Ablation Ion Source}
\shortauthors{K. Murray et~al.}

\title [mode = title]{Characterization of a Spatially Resolved Multi-Element Laser Ablation Ion Source}                      

\author[1]{K. Murray}[orcid=0000-0001-7065-4446]
\author[1]{C. Chambers}
\author[1]{D. Chen}
\author[1]{Z. Feng}
\author[1]{J. Fraser}
\fnmark[1]
\author[1]{Y. Ito}
\fnmark[2]
\author[1]{Y. Lan}
\author[4]{S. Mendez}
\fnmark[3]
\author[1]{M. Medina~Peregrina}
\author[1]{H. Rasiwala}
\author[1]{L. Richez}
\author[1]{N. Roy}
\author[1]{R. Simpson}
\fnmark[4]
\author[2,3]{J. Dilling}
\author[5]{W. Fairbank,~Jr.}
\author[2]{A.A. Kwiatkowski}
\author[1]{T. Brunner}[orcid=0000-0002-3131-8148]

\address[1]{McGill University, Montreal, Canada}
\address[2]{TRIUMF, Vancouver, Canada}
\address[3]{University of British Columbia, Vancouver, Canada}
\address[4]{Maastricht University, Maastricht, Netherlands}
\address[5]{Physics Department, Colorado State University, Fort Collins, Colorado, USA}
\fntext[fn1]{Now at Université de Montreal}
\fntext[fn3]{Now at ASRC JAEA, Ibaraki, Japan}
\fntext[fn3]{Now at University of Amsterdam}
\fntext[fn4]{Now at TRIUMF}

\begin{abstract}
A laser ablation ion source (LAS) is a powerful tool by which diverse species of ions can be produced for mass spectrometer calibration or surface study applications. It is necessary to frequently shift the laser position on the target to selectively ablate materials in a controlled manner, and to mitigate degradation of the target surface caused by ablation. An alternative to mounting the target onto a rotation wheel or $x-y$ translation stage, is to shift the laser spot position with a final reflection from a motorized kinematic mirror mount. Such a system has been developed, assembled and characterized with a two axis motorized mirror and various metal targets. In the system presented here, ions are ablated from the target surface and guided by a 90$^\circ$ quadrupole bender to a Faraday cup where the ion current is measured. Spatially resolved scans of the target are produced by actuating the mirror motors, thus moving the laser spot across the target, and performing synchronous measurements of the ion current to construct 2D images of a target surface which can be up to 50~mm in diameter. The spatial resolution of the system has been measured by scanning the interfaces between metals such as steel and niobium, where it was demonstrated that the LAS can selectively ablate an area of diameter $\approx$50~$\mu$m. This work informs the development of subsequent LAS systems, that are intended to serve as multi-element ion sources for commercial and custom-built time-of-flight mass spectrometers, or to selectively study surface specific regions of samples.
\end{abstract}


\maketitle

\section{Introduction}

The use of lasers in the production of ion beams, whether by resonance ionisation or ablation, has been demonstrated in many applications, see e.g. \cite{harkewicz1994solid,rencheng1997laser,gammino2002production,chichkov1996femtosecond,okamura2014laser,khalil2017tungsten, saquilayan2018production, munemoto2014development, twelker2014apparatus}. Ions may be produced in vacuum, or at higher pressures with different species of ambient gases \cite{elsied2018nanosecond}. Since the process of ablation is inherently destructive, it is typically necessary to shift the laser spot position on a target as the target surface degrades. Moreover, it may be desirable to selectively ablate specific areas of a multi-element target for applications such as mass spectrometer calibration. In high-precision mass-spectrometry applications, the positioning of a laser beam on the target surface is typically achieved by mounting the target onto a rotating wheel or $x-y$ translation stage \cite{KARNAKIS20067823,muller2009initial,riedo2013coupling,koestler2008high,green2017characterization,wiesendanger2017improved,cui2012depth,bauer2015heat}. In a recent development, the laser spot position was randomised by the movement of an external focusing lens which was mounted onto an $x-y$ translation stage \cite{tajima486offline}.  An alternative method, presented here, is to shift the laser spot position on the target with the final reflection into the vacuum chamber, using a high-precision motorized kinematic mirror mount. \newline 

The Laser Ablation Ion Source (LAS) presented in this work is being developed to provide ions with well known mass to calibrate a multi-reflection time of flight mass spectrometer \cite{murray2019design}. This development is part of the Ba-tagging development \cite{brunner2017searching} that is being pursued for a potential upgrade to the nEXO detector \cite{kharusi2018nexo}.\newline

The LAS uses a pulsed UV laser to produce ion bunches from a custom target in an Ultra-High Vacuum (UHV) system. For ablation to occur, an incident laser spot fluence of approximately $0.1-10$ J/cm$^2$ is required  \cite{ko2006nanosecond,jordan1995pulsed,chung2013nanosecond, HASHIDA2002862,cabalin1998experimental}. Target materials are chosen such that they span a large mass range for the purposes of mass-spectrometer calibration. The resulting ion bunches are guided by a quadru\-pole bender to a Faraday cup where the ion current is measured to characterize the performance of the setup. A 2D image of the target can be reproduced by rastering the ablation laser spot across the surface of the target and recording the ion current at each location. Since the ablation yield is material dependent \cite{baraldi2011dynamics}, the measured ion current changes for different materials, thus it is possible to distinguish physical features of the target. The resolution of physical features of the target allows for the establishment of a coordinate system on the target surface, which can be used to selectively ablate different materials. This work characterizes the spatial resolution with which the LAS may selectively ablate the surface of a target. 

\section{Experimental Setup}

The LAS consists of a vacuum chamber where the ions are produced and measured, attached to an optical breadboard where the laser beam optics are located. The vacuum chamber is a 6-way DN160 CF cross, which houses the ion-source assembly, 90$^\circ$ DC quadrupole deflector and a Faraday cup. UHV conditions are achieved by a turbomolecular pump attached to the bottom flange of the cross. The pressure is monitored with a cold cathode gauge, and is typically in the range of $1\times 10^{-7}$ to $1\times 10^{-8}$~mbar. A schematic drawing of the setup is shown in Fig. \ref{fig:las_assm}.\newline

\subsection{Laser Scanning}

\begin{figure*}
    \centering
    \includegraphics[width=0.85\textwidth,]{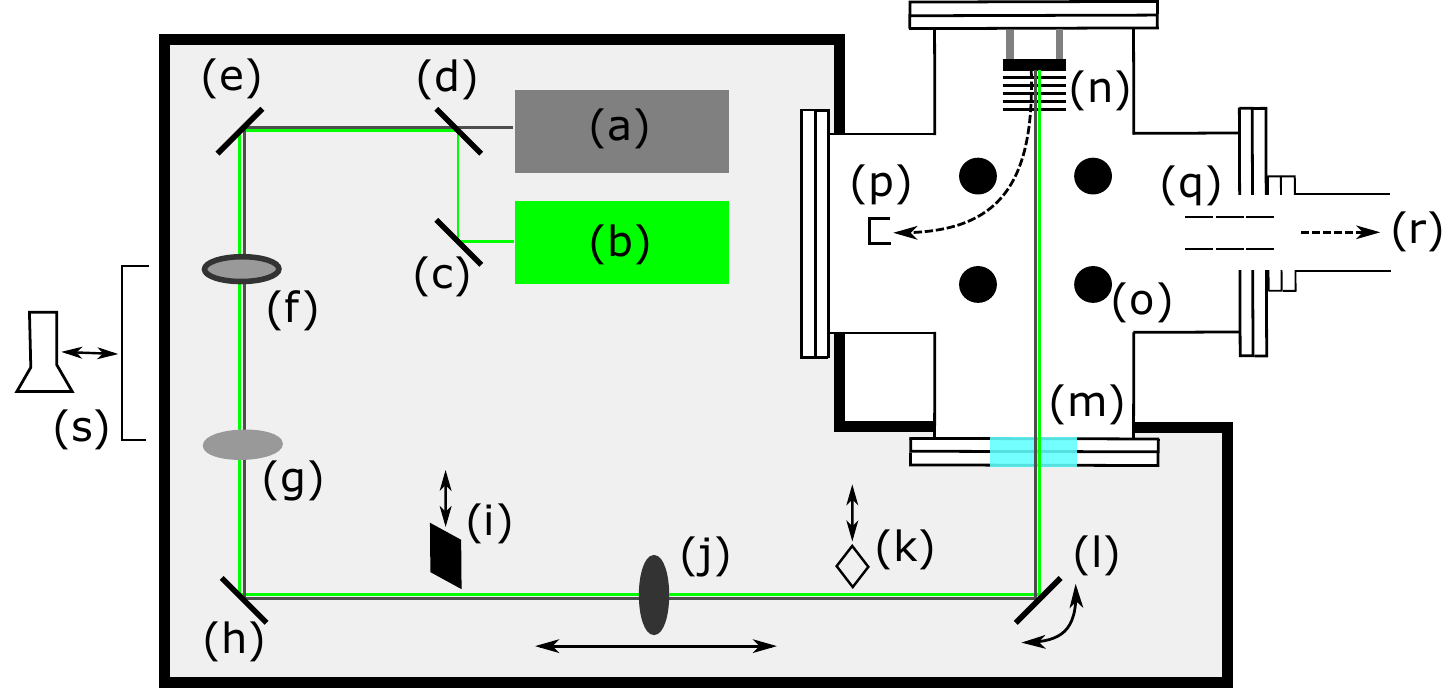} 
    \caption{Schematic of the LAS layout. The optical components guide a 349 nm Nd:YLF UV laser (a), and a 532 nm green laser (b), into a 6-way CF cross. These lasers are for ablation and alignment purposes, respectively, and a flip mirror (d) is used to toggle between both beams. The laser light is reflected by a series of UV enhanced aluminium mirrors (c)(e)(h), and passes through a pair of diverging (f) and converging (g) lenses. The beam is then focused onto a target with a $f$ = 750 mm UVFS plano-convex lens (j), which is mounted onto a motorized stage parallel to the optical axis. A knife-edge (i) and power meter (k) are used to measure the laser beam profile and energy. The beam is reflected onto the target by a UV-enhanced mirror mounted in a motorized kinematic mirror mount (l), which allows for precise manipulation of the laser spot on the target. The laser beam enters into the vacuum chamber through a UV transparent window (m), to strike the target (n). Once ions are produced from the target, they are accelerated by a series of ring electrodes. The ions then pass through a 90$^\circ$ quadrupole bender (o), to either a Faraday cup (p), or through an einzel lens (q) to another device. For measurements of ion transport efficiency, the diverging and converging lenses were replaced with a Thorlabs BE10-UVB beam expander (s).}
    \label{fig:las_assm}%
\end{figure*}

Ions are produced by a pulsed 349 nm neodymium-doped yttrium lithium fluoride (Nd:YLF) UV Spectra Physics Explorer 349 120. A series of mirrors and lenses focus the laser beam into the vacuum chamber. The UV laser can be pulsed with a variable frequency ranging from 1 Hz to 5 kHz, with a nominal energy of 120 $\mu$J per pulse, and a pulse width $< 5$ ns \cite{expmanual}. A visible laser is included in the setup to ease the alignment process.\newline 

The UV laser beam is focused onto the target by a plano-convex lens with a 750 mm focal length, which is mounted onto a motorized stage oriented parallel to the beam path, so that the location of the focal plane may be remotely adjusted to optimize ablation at the target surface. After passing through the focusing lens, the laser beam is reflected into the chamber and onto the target by a mirror on a computer-controlled motorized mount. Motorised mirror mounts are used to selectively ablate materials off of a fixed target (Fig. \ref{fig:las_assm} (l)). Two different motorized mirror mounts were used in this study. The first is a KS1-Z8 mount from Thorlabs with Z-812 series DC motor actuators. The second is a Physik Instrumente (PI) mount with N-472 linear actuators and E-871 servo controllers. The Thorlabs mount has a travel range of 12~mm, a minimum incremental movement of 0.2~$\mu$m, and a bidirectional repeatability of $<1.5$~$\mu$m. The PI mirror has a travel range of 13~mm, a minimum incremental movement of 0.05~$\mu$m, and a unidirectional repeatability of 0.2~$\mu$m.\newline

The mirror motors are controlled by a LabVIEW program that steps the laser spot in a rasterized grid across the target's surface, while recording the ion current measured at each point in the grid with the Faraday cup (Fig. \ref{fig:las_assm} (p)). In this manner, one and two dimensional scans of a target may be produced. 

\subsection{Ion Source and Ion Optics}

\begin{figure*}
\begin{subfigure}{.7\textwidth}
  \includegraphics[width=0.7\linewidth]{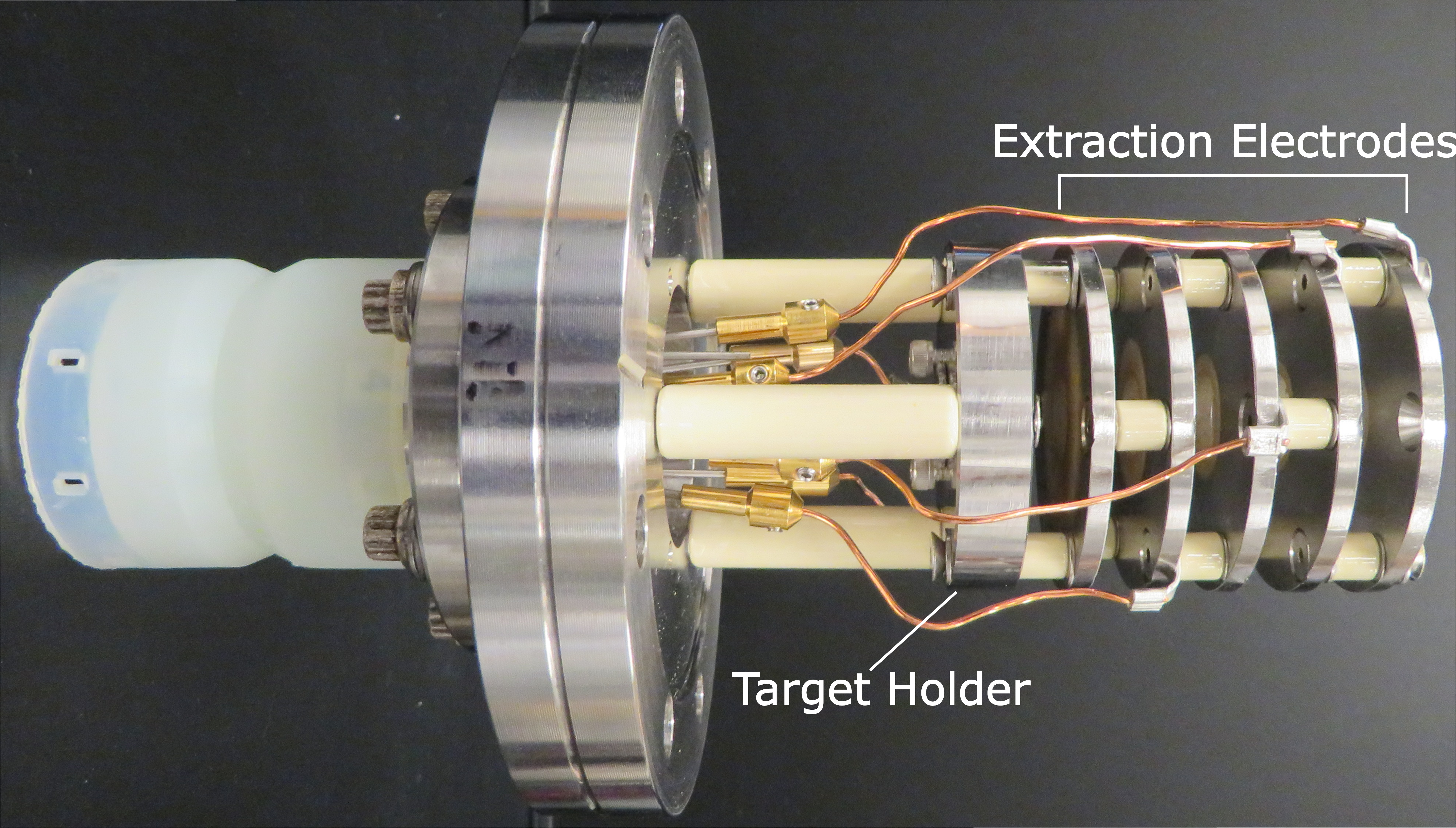}
\end{subfigure}%
\begin{subfigure}{.3\textwidth}
    \hspace{-0.7in}
  \includegraphics[width=0.96\linewidth]{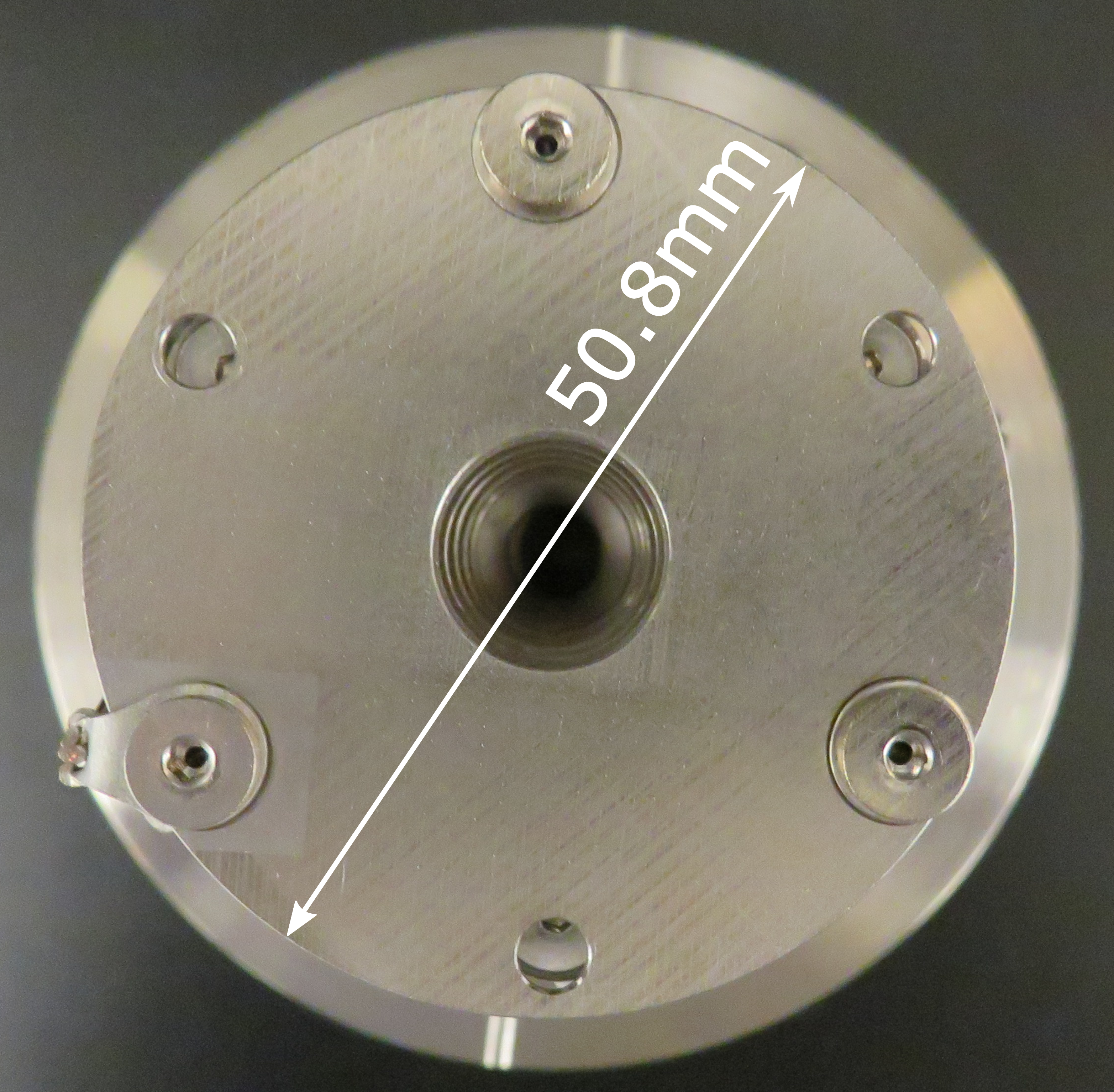}
\end{subfigure}
\caption{Photograph of the ion-source assembly with a view from the side (left) and from the front (right). The ion source consists of a steel target-holder, which is followed by 5 electrodes that create an accelerating DC field. The target-holder and electrodes have an outer-diameter of 50.8~mm, an inter-electrode spacing of 6.4~mm, and an inner diameter of 12.7~mm. The assembly is mounted onto a DN40 CF flange with ceramic standoffs.} 
\label{fig:source}
\end{figure*}

The ion-source assembly, shown in Fig. \ref{fig:source}, enables the mounting of custom targets in a stainless steel target-holder. Ions produced by ablation are accelerated by a series of five extraction electrodes, after which they are deflected by a 90$^{\circ}$ quadrupole bender to either a Faraday cup or another device such as a mass spectrometer. The 90$^{\circ}$ quadrupole bender is formed by 25.4~mm diameter stainless steel rods, which form a square with a center-to-center spacing of 59.1~mm between adjacent rods. The quadrupole bender's widely spaced electrodes allow surface scans over ranges as large as 50~mm in width.\newline

DC potentials are applied to the ion source, quadrupole bender and Faraday cup with Rohde and Schwarz HMC804 power supplies, which can output up to 32V per channel. In the current work, only the Faraday cup has been used since the mass spectrometer is still under development. The ion current produced by the ion source is typically measured with an Agilent 34465A digital multimeter connected to the Faraday cup. For direct measurement of the ion transport efficiency through the ion-source assembly and the quadru\-pole bender, the ion currents leaving the target and arriving at the Faraday cup were measured simultaneously using two Keithley 6485 picoammeters. For measurement of the ion transport efficiency, the LAS had been upgraded by replacing the diverging and converging lenses with a Thorlabs BE10-UVB beam expander to create a smaller laser diameter on the target. A cylindrical Cu only target was used to investigate the homogeneity of the ion transport efficiency across the target surface with a 2D scan.\newline

\section{Target Scanning in 2D}

The actuated mirror moves the laser beam in two directions across the target surface. The directions of movement are to first order perpendicular to each other. The system is calibrated so that the distance actuated by the mirror stages can be converted to distance travelled on the target surface. The inner diameter of the stainless steel target holder is used as a reference to construct the target coordinate system from a 2D scan of the target. This coordinate system can be described simply with the following transformation

\begin{equation}
(x',y') = (S_x x ,S_y y ), 
\label{eq:proj_fac}
\end{equation}
where ($x$,$y$) are the coordinates of the mirror stages, ($S_x$,$S_y$) are scaling factors, and ($x'$,$y'$) are the target coordinates.\newline

\begin{figure}[pos=htp]
\centering
\includegraphics[width=0.6\linewidth]{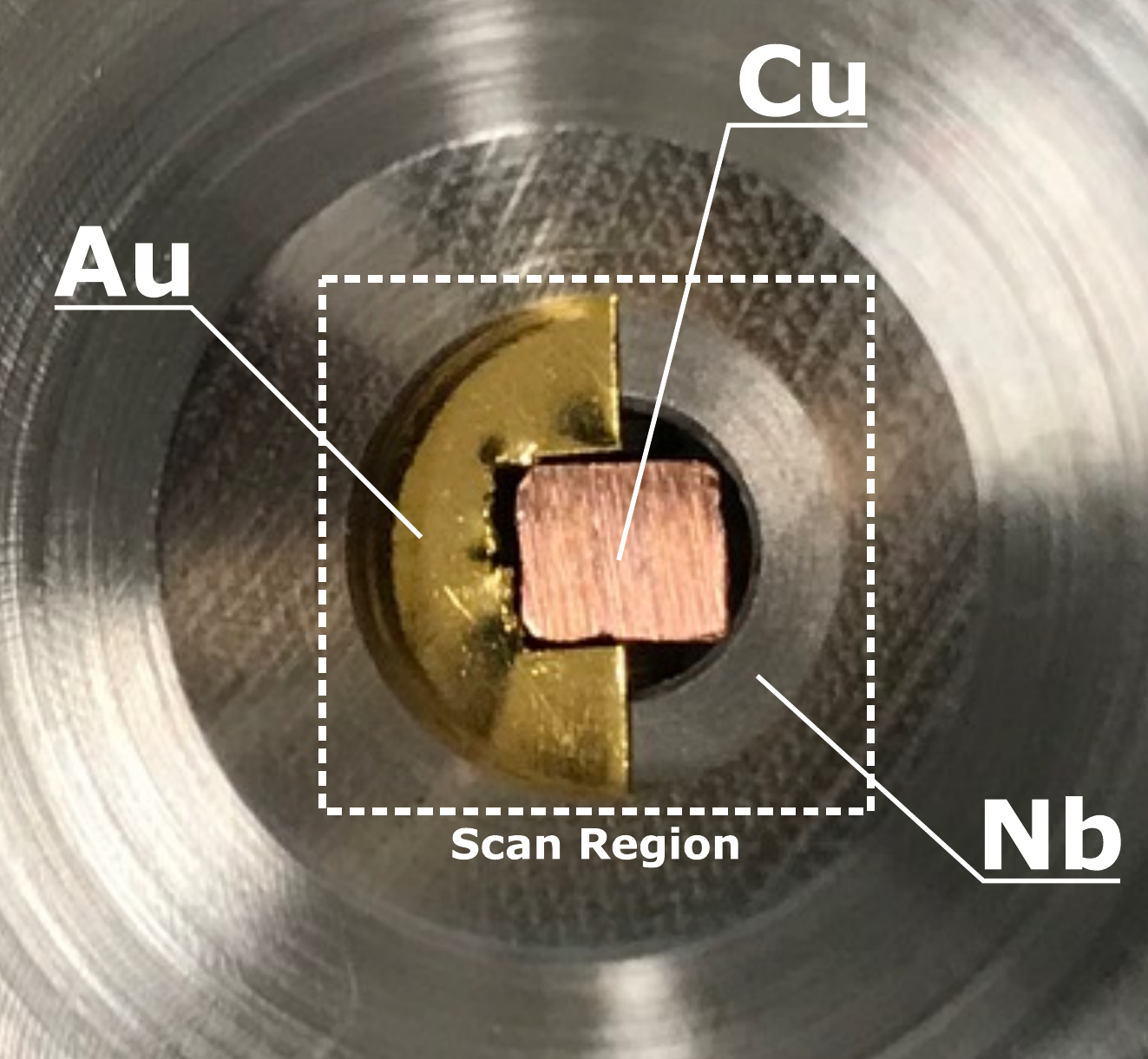}
\caption{A photograph of the Au-Nb-Cu target in a steel holder that forms the base of the ion source. The inner diameter of the steel holder is 7.8(1)~mm, and this is used as a reference dimension when calibrating the laser spot motion on the target.} 
\label{fig:aucunb}
\end{figure}

A multi-element target consisting of gold, niobium and copper (Au-Nb-Cu) was scanned in 2D by the LAS. Scans of the Au-Nb-Cu target are used here to demonstrate a measurement of the scaling factors described by Eq. (\ref{eq:proj_fac}). The Au-Nb-Cu target is formed by a hollow niobium cylinder, with a rectangular copper rod in its centre held in place by stainless steel set screws. A 0.1~mm thick piece of gold foil covers one half of the niobium face. The target is shown mounted into the target holder in Fig. \ref{fig:aucunb}. These metals have been chosen such that ions from the target span a significant mass range, for the benefit of mass-spectrometer calibration.\newline

\subsection{Determination of Scaling Factors}
\begin{figure}[pos=htp]
    \centering
    \begin{subfigure}{.55\textwidth}
      \centering
      \includegraphics[width=0.9\linewidth]{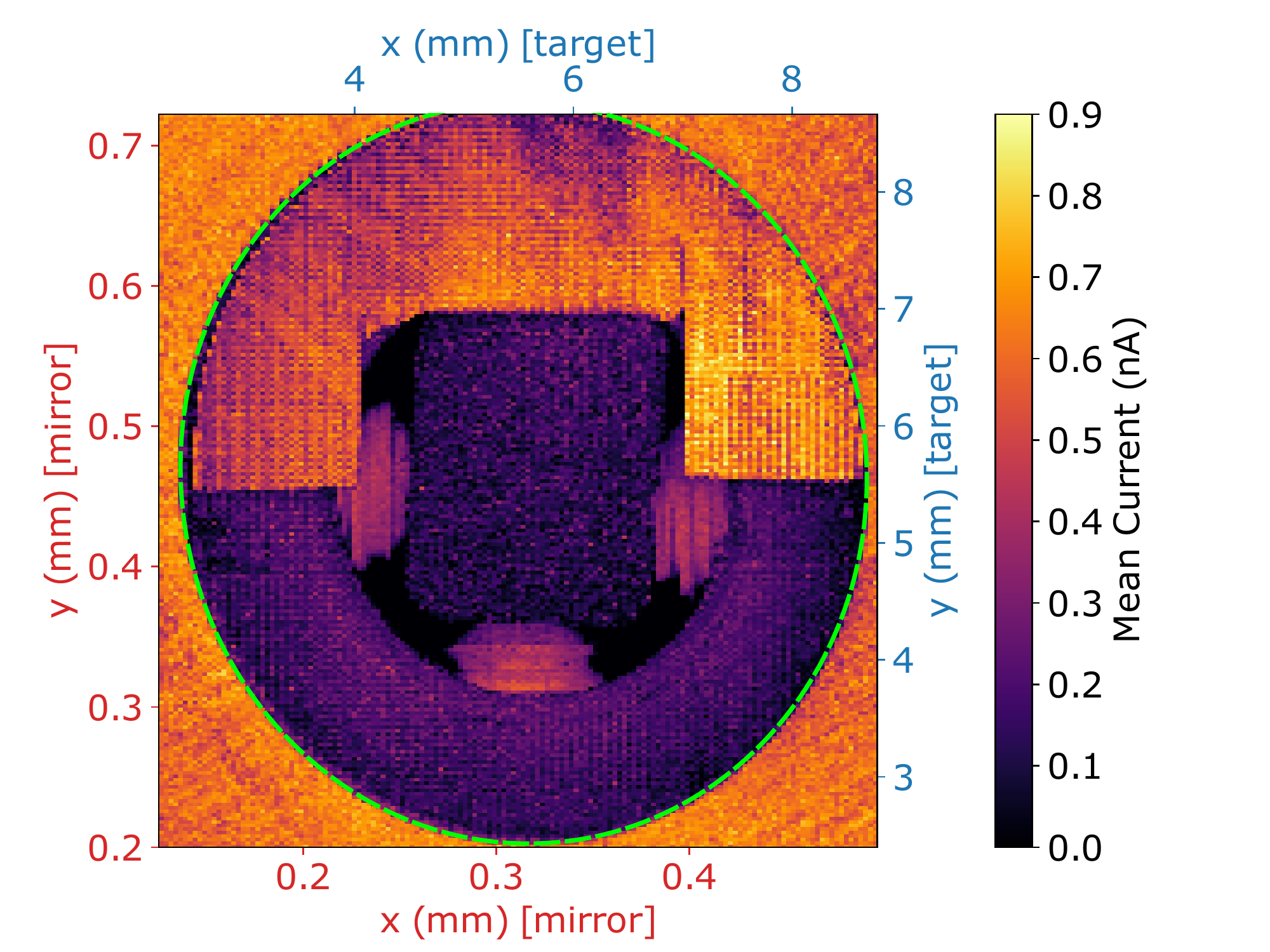}
      \label{fig:2DPI}
    \end{subfigure}
    \hspace*{-0.12in}
    \begin{subfigure}{.53\textwidth}
      \centering
      \includegraphics[width=0.9\linewidth]{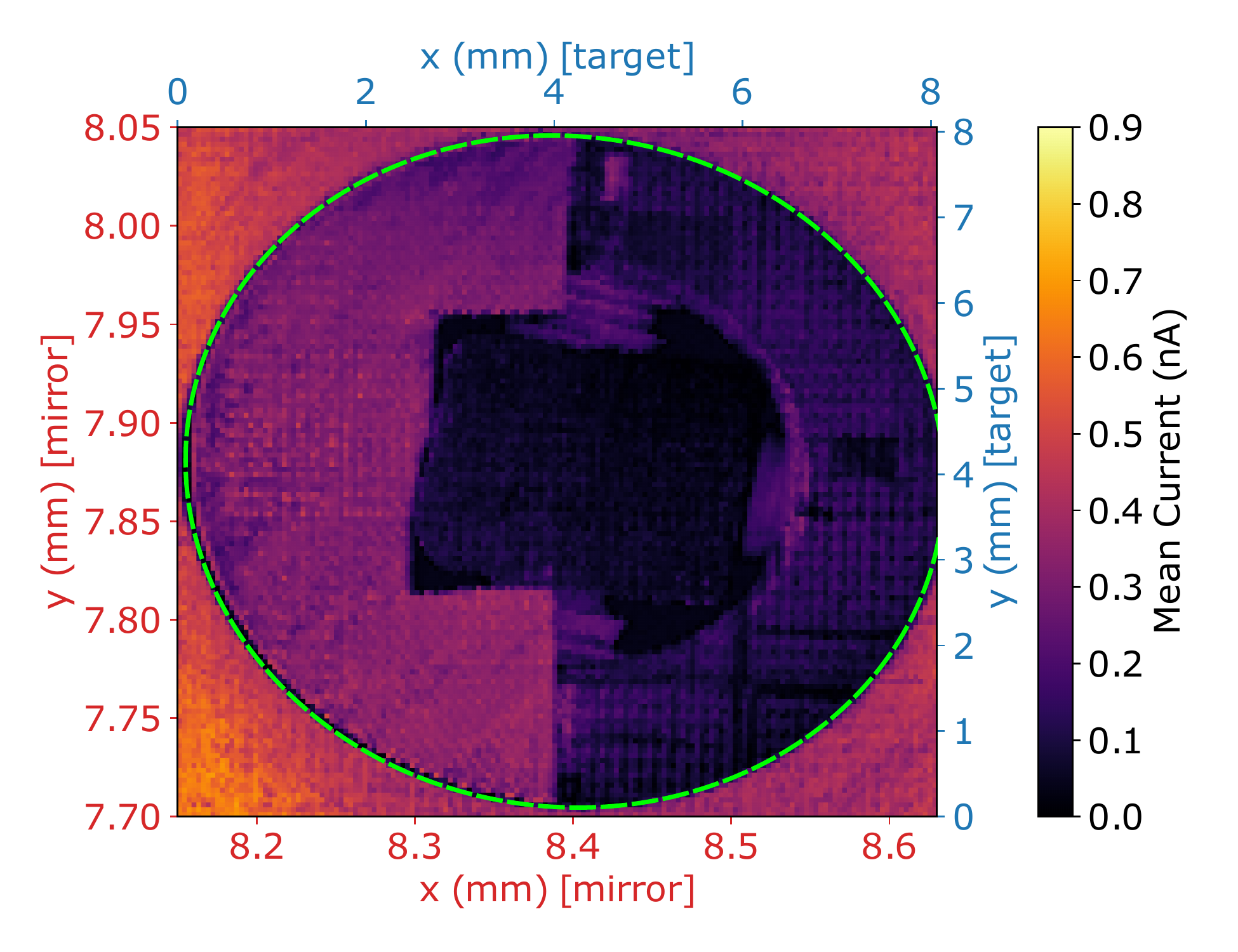}
      \label{fig:2DThor}
    \end{subfigure}
    \caption{A 2D reconstruction of the target surface, using the PI (top) and Thorlabs (bottom) kinematic mirror mounts to adjust the laser spot position, where the colorbar indicates the ion current measured by the Faraday cup. The scaling factors used to convert to target coordinates are obtained by fitting an ellipse to the inner diameter of the steel holder, which is marked in each image by a green dashed line, and comparing this to the physically measured value of 7.8(1)~mm. The target was physically rotated by 90$^\circ$ counterclockwise between top and bottom scan image and the gold foil had been replaced.}
    \label{fig:2Dscan}%
\end{figure}
The scaling factors were calculated for each mirror, by creating a 2D scan of the target and fitting an ellipse to the diameter of the target holder. Two examples are shown in Fig. \ref{fig:2Dscan} for the PI and Thorlabs mirror mounts on the top and bottom respectively. The ellipse was fit to a total of 30 points which were selected manually at the interface of the target and the target holder. The diameter of the target holder was measured as $D_{\text{TH}}$ = 7.8(1)~mm, $S_x$ and $S_y$ were calculated as the ratio of $D_{\text{TH}}$ with the best-fit values for the semi-minor and semi-major axes respectively. The measured scaling factors for both mirrors are summarized in Table \ref{tab:scalings}, where the smallest step size on the target surface for each mirror is also presented. As discussed in Section \ref{sptl res}, the spatial resolution is limited by the diameter of the laser spot on the target, and not the smallest step size. Thus, both mirror mounts are able to reconstruct the physical features of the target with similar performance. Small differences in the mean ion current between the two scans are not caused by the use of different motorized mirror mounts, but by slight differences in the optical setup, as the scans presented here were taken at different stages of the development of the technique. Additionally, the surface of the target was scanned repeatedly over the development period, and differences in the amount of damage accumulated on the target surface could result in differences in the measured ion current. \newline

\begin{figure}[pos=htp]
\centering
\begin{subfigure}{.5\textwidth}
  \includegraphics[width=\linewidth]{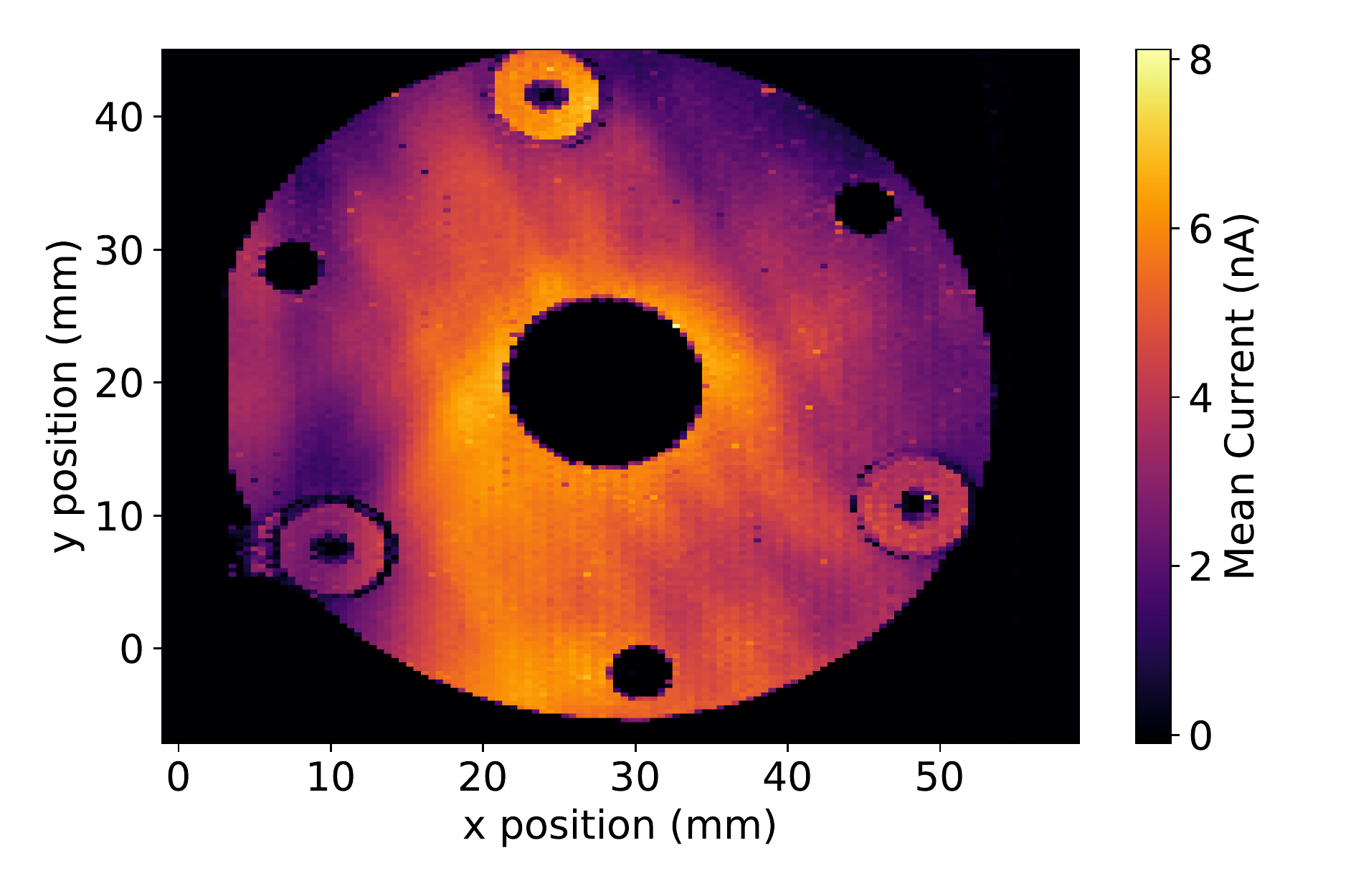}
\end{subfigure}\newline
\begin{subfigure}{.45\textwidth}
  \includegraphics[width=\linewidth]{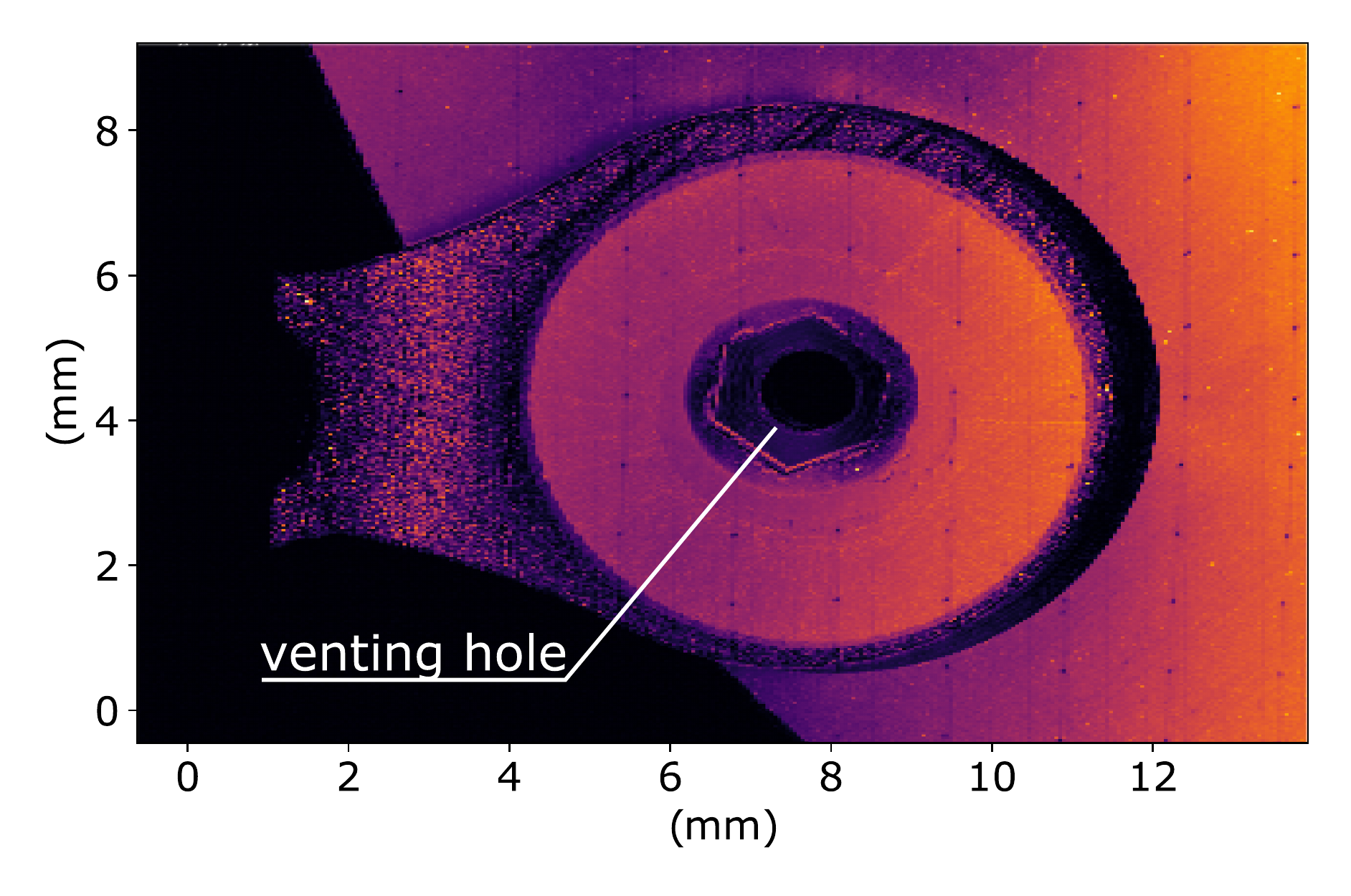}
\end{subfigure}
\caption{ (top) A 2D scan of the final electrode of the ion-source assembly, where the colorbar indicates the ion current measured by the Faraday cup. A mirror step size of 25~$\mu$m in both the $x$ and $y$ directions was used, which corresponds to step size on the target of 475(8)~$\mu$m  and 320(5)~$\mu$m respectively. (bottom) A detailed scan of a socket-head screw used to secure the electrode, where the mirror step size is a factor of 10 smaller. The dark circle is the hole of the center-vented screw.} 
\label{fig:plat5}
\end{figure}

\begin{table}[pos=htp]
\centering
\begin{tabular}{l | c | c | c | c}
\hline
&\multicolumn{2}{c}{Scaling Factor} & \multicolumn{2}{c}{Smallest Step Size}\\
\hline
Mirror Mount & $S_x$ & $S_y$& $x_{\text{min}}$($\mu$m) & $y_{\text{min}}$($\mu$m)\\
\hline
TL KS1-Z8 & 19.2(3) & 13.9(2)& 3.84(6) & 2.78(4) \\
PI N472 & 21.8(3) & 14.9(2)& 1.09(2) & 0.75(1) \\
\hline
\end{tabular}
\caption{The measured scaling factors for $x$ and $y$  for the PI and Thorlabs actuated mirrors, obtained through the fit of an ellipse as demonstrated in Fig. \ref{fig:2Dscan}.}
\label{tab:scalings}
\end{table}

The large spacing between the quadrupole electrodes allows for scanning over ranges as large as 50~mm. To demonstrate this, the focusing lens, which is mounted onto a motorized stage, was adjusted so that the laser beam was focused on the surface of the final accelerating electrode. A 2D scan of the full electrode is shown on top in Fig. \ref{fig:plat5}. By fitting an ellipse to the inner diameter, the scaling factors were measured as $S_x$=19.0(3) and $S_y$=12.8(2) for the PI mirror. The scan of the full electrode was taken over the period of a day with step sizes of 475(8)~$\mu$m and 320(5)~$\mu$m in $x$ and $y$ respectively. A detailed scan of one of the bolts securing the electrode is shown at the bottom of Fig. \ref{fig:plat5} with a 10-fold smaller step size. The scaling factors were found to remain consistent between measurements, but the system needs to be re-calibrated if there are changes to the optical setup. 

\subsection{Measurement of Laser Spot Step Size}

An optical microscope was used to directly measure the distance between ablated spots, to confirm that the scaling factors accurately predict the laser spot step size on the target surface. A thin Cu foil was scanned over the full target range using the PI mirror mount, with a laser beam energy of 18.1(2) $\mu$J per pulse at 500 Hz and a fixed step size of 20~$\mu$m in mirror coordinates in both the $x$ and $y$ directions. Since the laser emitted pulses while the spot was being moved on the target surface, there were ablation craters as well as tracks that connect them. To include scans at a different laser beam energy, two opposite quarters of the target were scanned a second time with a higher laser beam energy of 28.8(3) $\mu$J and 36.1(6) $\mu$J per pulse. It was found that the size of craters and tracks created by 28.8(3) $\mu$J and 36.1(6) $\mu$J laser pulses were indistinguishable under the microscope, but distinguishable from craters and tracks formed by 18.1(2) $\mu$J pulses.\newline

\begin{figure}[pos=htp]
    \centering
    \includegraphics[width=0.45\textwidth,]{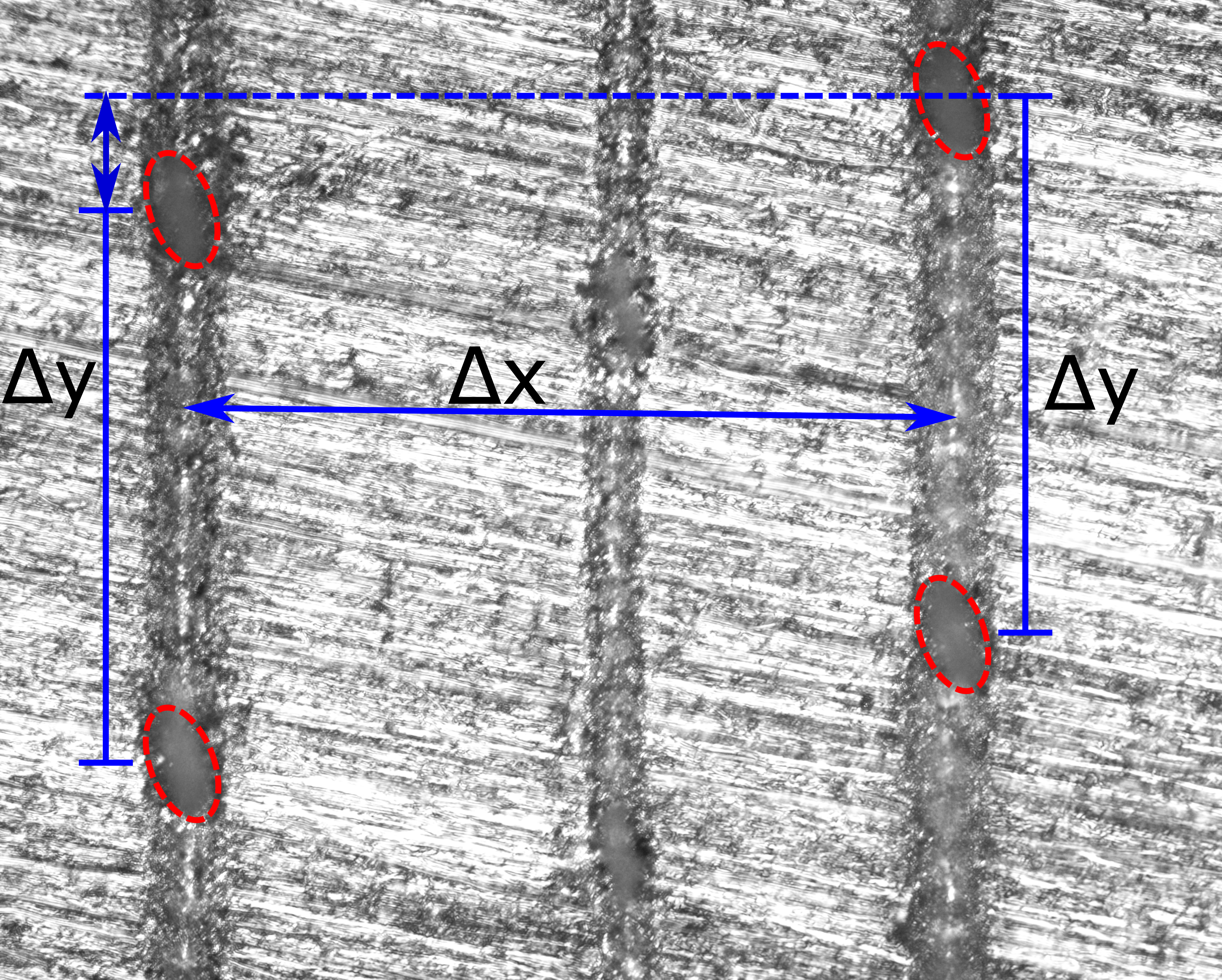} 
    \caption{Image of ablation craters and tracks on the surface of Cu foil recorded with an optical microscope at $\times$20 magnification. The distances in $x$ and $y$ between adjacent ablation craters, denoted by red-dashed lines, were measured as $\Delta x$ = 441.6(6)~$\mu$m and $\Delta y$ = 305.1(9)~$\mu$m respectively. The thinner track in the center was formed by a different scan with a laser beam energy of 18.1(2) $\mu$J per pulse and is not relevant to the presented step-size measurements.}
    \label{fig:abl_cu}%
\end{figure}

The average $x$ and $y$ distances between the centre positions of 4 elliptical ablation craters, which are indicated with red dashed lines in Fig. \ref{fig:abl_cu}, were measured at $\times$20 magnification with an optical microscope. The ablation craters are elliptical in shape, with the average of the four major and minor axes measured as 68(5)~$\mu$m and 39(2)~$\mu$m respectively. The distance in $x$ between ablation points was measured as $\Delta x$ = 441.6(6)~$\mu$m, and the distance in $y$ was measured as $\Delta y$ = 305.1(9)~$\mu$m. This measured distance is in agreement with the predicted distances travelled in $x$ and $y$ of 438(6)~$\mu$m and 302(4)~$\mu$m, respectively, calculated using the scaling factors $S_x$ = 21.9(3) and $S_y$ = 15.1(2). It was found that the first step taken in $y$, after the mirror had reversed the scanning direction, was shorter than the subsequent steps in $y$. Thus there is a shift in $y$ between adjacent scan lines. For reference, the difference in the $y$ position of adjacent craters was measured as 66(2)~$\mu$m.\newline

\section{Spatial Resolution}
\label{sptl res}

The spatial resolution at which surface features can be resolved with the LAS system is limited by either the diameter of the ablated crater, or the smallest step size that the laser beam can move on the target surface. The diameter of the ablated crater can be measured directly with an optical microscope, for comparison with the smallest achievable steps calculated in Table \ref{tab:scalings}. Alternatively, the diameter of the ablated crater can be estimated theoretically with the following method.   

\subsection{Ablation Diameter}
The diameter of the ablated crater is related to the focused $1/e^2$ beam diameter, $D_f$, which can be calculated from a measurement of the laser beam width before the focusing lens, using the relationship \cite{self1983focusing,siegman1998maybe}
\begin{equation}
D_f = \frac{4 M^2 \lambda f}{\pi D},
\label{eq:DvF}
\end{equation}
where $f$ is the lens' focal length, $\lambda$ the laser wavelength, $D$ is the $1/e^2$ diameter of the beam as it enters the lens and $M^2$ is the beam quality parameter. It is important to note that material is only ablated where the laser spot fluence is higher than the ablation threshold of the material being ablated. The relationship between the ablated crater diameter, $D_a$, and the threshold fluence, $F_{th}$, is given by \cite{liu1982simple,semerok2001femtosecond,stafe2013pulsed}
\begin{equation}
D^2_a = \frac{D_f^2}{2}ln\left(\frac{F_0}{F_{th}}\right),
\label{eq:DavDf}
\end{equation}
where the peak fluence, $F_0$, is 
\begin{equation}
F_0= \frac{8 E_T}{\pi D_f^2}
\label{eq:peakF}
\end{equation}
and $E_T$ is the total beam energy per pulse. It is also important to note that the absorptivity of the irradiated zone changes after multiple pulses. The threshold fluence after $N$ pulses, $F_{th}(N)$ is related to the single-shot threshold fluence by a power law \cite{jee1988laser}
\begin{equation}
F_{th}(N) = F_{th}(1)N^{\zeta-1},
\label{eq:inc}
\end{equation}
where $\zeta$ is known as the incubation coefficient, and $F_{th}(1)$ is the fluence threshold of the first laser shot on a fresh target. Since the threshold fluence changes, $D_a$ will vary depending on the number of times a particular spot has been ablated, and damage accumulation will affect the spatial resolution for repeated pulses. \newline

The profile of the laser beam before focusing was measured by mounting a razor blade on a micrometer stage and moving the blade in steps of 0.25~mm across the beam, while measuring the average laser beam energy that passes the blade for 10 s at each step. The laser beam energy was measured after the focusing lens and before reflection by the motorized mirror. The resulting profile was fit with an integrated Gaussian function, from which the $1/e^2$ diameter $D$ = 3.9(1)~mm was extracted.\newline 

The $M^2$ beam quality parameter was specified for the laser as $M^2<1.3$ \cite{expmanual}. The $M^2$ parameter was also measured, by deflecting the laser beam after focusing away from the motorized mirror mount, and taking knife-edge measurements of the beam intensity profile near and around the focus, as outlined in \cite{siegman1998maybe}. The beam quality was measured as $M^2=1.32(7)$, which agrees with the laser specifications. Thus $D_f$ was calculated using Eq. (\ref{eq:DvF}), with $f = 750(8)$~mm, $\lambda = 349$~nm, and $M^2=1.32(7)$ as $D_f$ = 113(7)~$\mu$m, with a peak fluence of 0.9~J/cm$^2$. The focus of the laser beam on the target surface was optimized prior to this profile measurement by adjusting the position of the focusing lens, and maximizing the ion current produced from the steel target holder.\newline

To calculate the diameter of the ablated crater with Eq. (\ref{eq:DavDf}), $F_{th}(1)$ and $\zeta$ must be known. The incubation parameter $\zeta$ is sensitive to the pulse duration, and will typically have a value between 0.8-0.9 for femtosecond and nanosecond pulses respectively \cite{raciukaitis2008accumulation}. However, the single-shot fluence threshold $F_{th}(1)$ can vary by orders of magnitude depending on the wavelength \cite{hutson2009interplay,natoli2007influence,carr2003wavelength}, pulse duration and repetition rate \cite{brygo2006laser}, material composition and beam diameter \cite{naghilou2017femto,armbruster2017spot}. Thus it is difficult to choose a value for $F_{th}(1)$ when calculating the ablated crater diameter for this experiment. 

\subsection{Scanning Metal Junctions}
The metals of the multi-metal targets can be distinguished by the amplitude of emitted ion current produced at constant average laser beam power. This fact is utilized to perform a measurement of the spatial resolution for the setup. When the laser spot was scanned across the junction of two metals, the ion current appeared as a smeared step function. The width of this transition, $D_t$, was measured by fitting an integrated Gaussian function to the resulting distribution, from which the spatial resolution was extracted as the $1/e^2$ diameter. Both motorized mirror mounts were capable of repeatable increments smaller than $4$~$\mu$m on the target surface, which is significantly smaller than the focused laser spot diameter on the target. Thus the resolution was dominated by the laser spot diameter on the target, and both mirror actuator types were found to have similar spatial resolutions.\newline

\begin{figure}[pos=h]
    \centering
    \begin{subfigure}{.5\textwidth}
      \centering
      \includegraphics[width=0.9\linewidth]{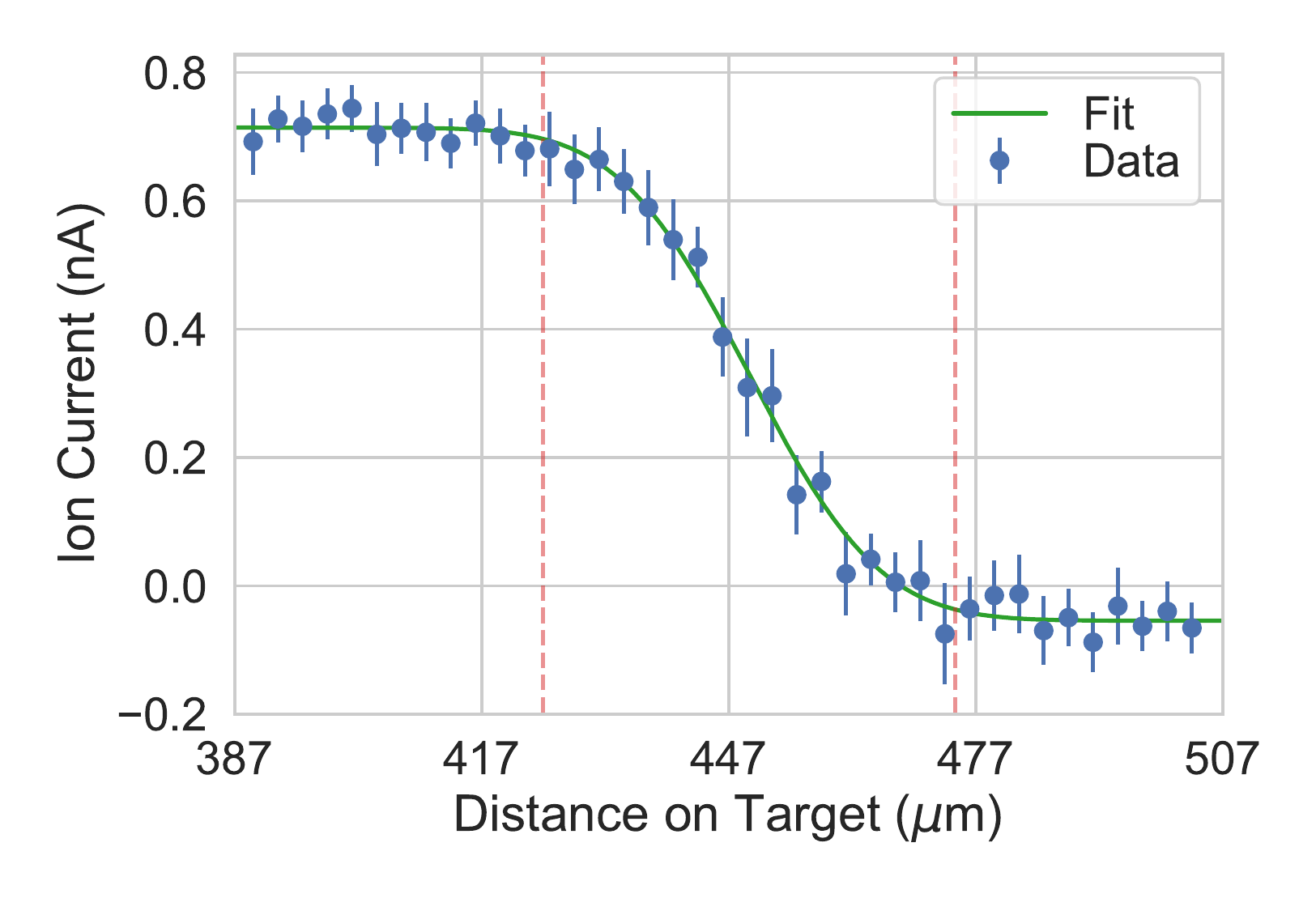}
      \label{fig:steel_niob}
    \end{subfigure}\newline
    \begin{subfigure}{.5\textwidth}
      \centering
      \includegraphics[width=0.9\linewidth]{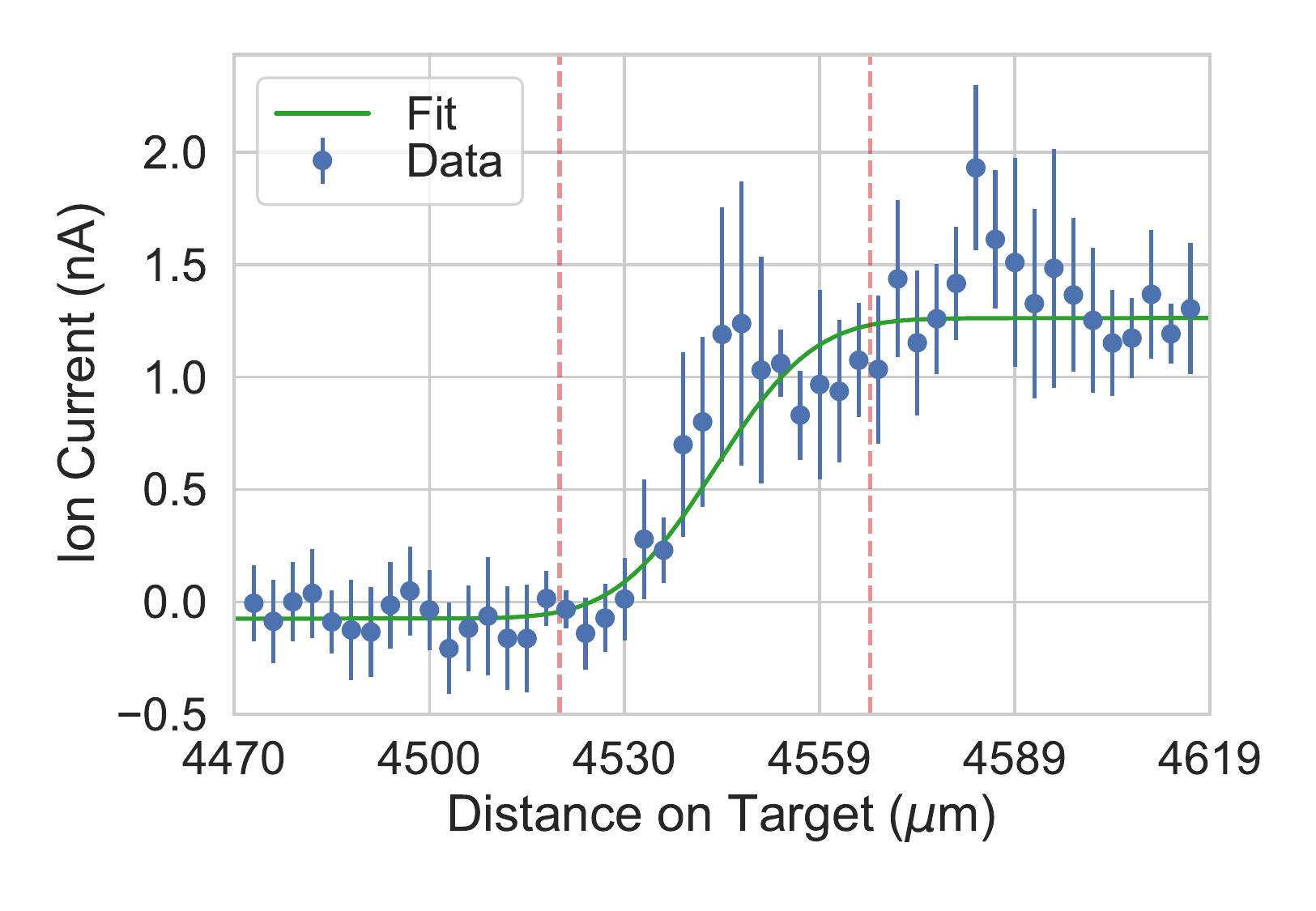}
      \label{fig:gold_silicon}
    \end{subfigure}
    \caption{A representative scan of the transition between steel and niobium (top), and silicon and gold foil (bottom). The profiles are fit with integrated gaussian functions from which the $1/e^2$ diameters are extracted as 50(3)~$\mu$m and 47(7)~$\mu$m respectively.}
    \label{fig:currscan}%
\end{figure}

The spatial resolution was measured with the Au-Nb-Cu target as well as an Al-Si-Au target. The Al-Si-Au target consisted of a 0.4~mm thick piece of aluminum foil and 0.1~mm thick gold foil on top of a silicon wafer, such that the upper portion of the face is covered with gold and the lower portion is covered by aluminum, with a 2~mm gap of silicon in between. This target provides straight edges of gold and aluminum that span the diameter of the target, allowing for more scans of metal transitions to be made before needing to replace the target. For the experiments presented here, the ion current was averaged over a time window of 0.25 s for each scan position, and the standard deviation was calculated and is presented as the statistical uncertainty. The laser repetition rate was fixed at 1000 Hz and measurements of the laser beam energy were taken at the location just before the beam reflects from the motorized mirror. The following measurements were performed with the PI mirror mount by scanning targets in the $y$ direction, thus the scaling factor $S_y$=14.9(2) is used to calculate the results. The junction between the steel target-holder and the niobium component of the Au-Nb-Cu target was scanned, and the resulting profile fit with an integrated Gaussian function. The $1/e^2$ full width of the transition was extracted as $D_t$ = 50(3)~$\mu$m, with the laser beam energy per pulse measured as 33(1) $\mu J$. The Al-Si-Au target was then installed such that the aluminum and gold edges were perpendicular to the $y$-direction. The junction of silicon and gold foil was scanned, and the width extracted as $D_t$ = 47(7)~$\mu$m, with the laser beam energy measured as 42(1) $\mu J$. These scans are displayed on the top and bottom of Fig. \ref{fig:currscan} for stainless steel-niobium and silicon-gold foil interfaces, respectively. Under the described experimental conditions no ablation was observed on the Al target. \newline

Using eqs. (\ref{eq:DavDf}-\ref{eq:inc}), the threshold fluence for stainless steel was estimated. With $D_a = 50(3)$~$\mu$m, $E_T = 33(1)$~$\mu J$ and $D_f = 113(7)$~$\mu$m, the single shot threshold fluence could be expressed as $F_{th}(1)=\nicefrac{0.4}{N^{\zeta - 1}}$~J/cm$^2$. Using the laser repetition rate, $f_r$, the measurement time, $t_m$, the scan step size, $\Delta s$, and the beam diameter, the number of overlapping pulses was estimated as $N\approx f_r \times t_m \times D_f/\Delta s \approx 1000 \times 0.25 \times 113/2 \approx 14000$. Thus the single-shot threshold fluence was calculated as $F_{th}(1)\sim 1.2$~J/cm$^2$, assuming an incubation coefficient of $\zeta = 0.9$ \cite{raciukaitis2008accumulation}. It could be argued that the choice of $D_f$ as the overlap length to be divided by $\Delta s$ is somewhat arbitrary, and that $N$ falls within a range of values. However, the value of $F_{th}(1)$ is not strongly dependent on $N$, provided that $N \sim \mathcal{O} (1000)$ pulses. For example, if $D_a$ had been chosen as the overlap length instead, then $N \approx 6250$ and $F_{th}(1)\sim 1.0$~J/cm$^2$. Taking a range of values for $N$ from 6250 to 14000 results in an estimate for the single-shot threshold fluence of $F_{th}(1)\sim 1.0-1.2$~J/cm$^2$. This result is comparable to the results of a study where, for a comparable beam diameter, the threshold fluence after 100 pulses was found to be $F_{th}(100)\sim 0.8$~J/cm$^2$ for stainless steel \cite{naghilou2017femto}. Assuming the same incubation coefficient, this corresponds to a single-shot threshold fluence of $F_{th}(1) \sim 1.3$~J/cm$^2$.

\section{Ion Transport Efficiency}

The ion transport efficiency, $\text{T}_{\text{eff}}$, is measured as the ratio of the ion current arriving at the Faraday cup, $\text{I}_{\text{FC}}$, to the ion current leaving the target holder, $\text{I}_{\text{TH}}$, \newline
\begin{equation}
    \text{T}_{\text{eff}} = \frac{\text{I}_{\text{FC}}}{\text{I}_{\text{TH}}}.
    \label{eq:eff}
\end{equation}

However, the process of laser ablation does not only release singly charged ions, but electrons, neutrals and multiply-charged ions as well \cite{farid2014emission}. Thus, Eq. (\ref{eq:eff}) is only an accurate measurement of the ion transport efficiency if the electrons are re-captured by the target-holder, and the distribution of ionisation states in the ablated material is dominated by $+1$. To prevent electrons escaping the target holder, a positive DC bias of 32V was applied to the target holder. A larger DC bias than what was used for the previous experiments was applied to the quadrupole bender electrodes to increase the bending efficiency, with potentials of -200~V and 40~V applied to the quadrupole electrodes nearest to the Faraday cup.\newline

The laser beam energy and DC biases applied to the ion source were adjusted to maximise the transport efficiency, while ensuring that the potential gradient between the target holder and the first extraction electrode was sufficient to re-capture electrons on the target holder. Replacing the diverging and converging lenses with the Thorlabs beam expander increased the diameter of laser beam as it enters the focusing lens from 3.9$(1)$~mm to 9.6$(1)$~mm, thus reducing the calculated spot diameter on the target to $D_f$ = $45(1)$~$\mu$m. The laser repetition rate was fixed at 500 Hz for the experiment and the laser beam energy measured as 18.1(1) $\mu J$ per pulse, which calculates to a peak fluence of 2.3~J/cm$^2$. The laser beam energy was measured by placing the energy meter at the target's location, but without the target in place. The Cu target was scanned with a step size of 109(2)~$\mu$m in $x$ and 75(1)~$\mu$m in $y$. The current was measured simultaneously at the target holder and Faraday cup for 8~s at each point of the scan.\newline

\begin{figure}[pos=htp]
    \centering
    \includegraphics[width=0.45\textwidth,]{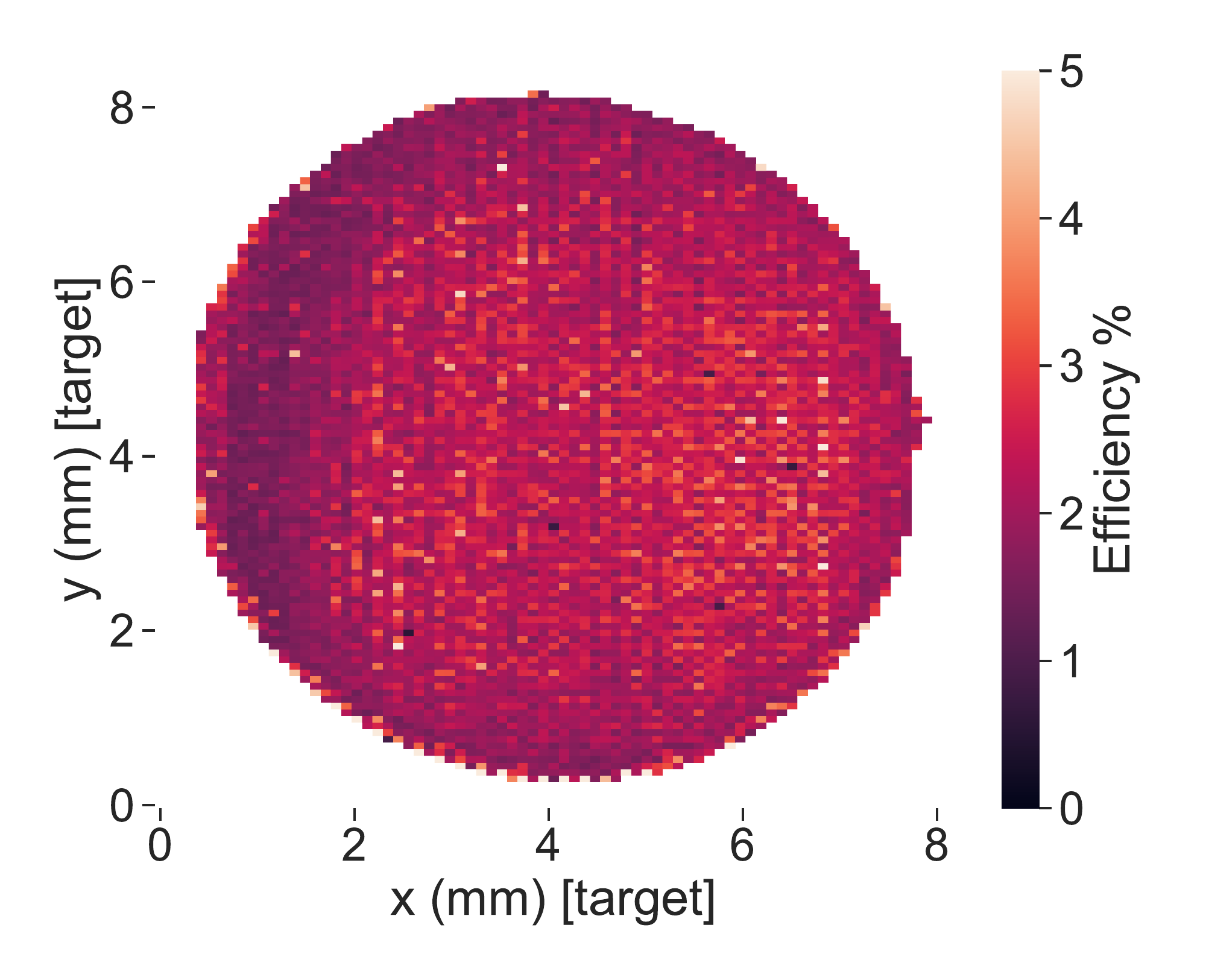} 
    \caption{A 2D scan of the ion transport efficiency, calculated with Eq. (\ref{eq:eff}), across the surface of a solid Cu target, with a step size of 109(2)~$\mu$m in $x$ and 75(1)~$\mu$m in $y$.}
    \label{fig:eff_plot}%
\end{figure}

For 5741 scan points across the surface of the Cu target, the average current measured leaving the target holder was 121(17) nA with the uncertainty taken as the standard deviation. The average current measured arriving at the Faraday cup for the same scan points was 2.6(2) nA. From Eq. (\ref{eq:eff}), the average ion transport efficiency across the target surface was calculated to be 2.2(7)\%. A plot of the calculated efficiency for each scan point is shown in Fig. \ref{fig:eff_plot}. The efficiency is fairly homogeneous across the surface of the target, with the left side slightly less efficient than the right. This small gradient could be caused by the difference in proximity to the bending quadrupole electrodes.\newline

For the ablation of Cu with a 308 nm laser spot at a fluence of 3 J/cm$^2$, a study has shown that the charge state distribution has a Cu$^{1+}$/Cu$^{2+}$ ratio of 15.2 \cite{torrisi2003comparison}. There, the average kinetic energy of ablated Cu ions was $\mathcal{O} \sim(100)$~eV and the angular distribution of ablated yield was found to have a Full Width at Half Maximum (FWHM) of 36$^{\circ}$. This study had similar ablation conditions to the system presented in this work, thus singly charged Cu ions are expected to dominate the ablated ions, and ions are expected to have similar angular and kinetic energy distributions. In the system presented here, the aperture of the final electrode subtends an angle of $\sim 15^{\circ}$. Assuming a Guassian angular distribution, $\sim$65\% of ablated ions are expected to be lost by angular acceptance due to collision with the extraction electrodes, primarily the first extraction electrode.\newline

Further ion loss is expected to result from inefficient bending of the quadrupole bender. The ion transport efficiency has been investigated using SIMION [45] simulations based on published ion distributions \cite{torrisi2003comparison,jordan1995pulsed,baraldi2011dynamics}, and gave results comparable to the measured ion transport efficiency. Further simulations were conducted with ions leaving normal to the target surface with energies from 0 eV to 80 eV, with an increment of 1 eV between consecutive ions. Simulation results indicated that the widely spaced electrodes of the quadrupole bender, with the same potentials applied to the quadrupole electrodes as in the experiment, caused the bender to effectively act as a kinetic energy filter. In simulations, the quadrupole bender was unable to sufficiently bend Cu$^{1+}$ ions with a kinetic energy greater than $\sim$20~eV towards the Faraday Cup for collection. This cut-off in kinetic energy was independent of the ion's origin on the target surface within a range of $\sim2$~eV, and was significantly lower than the expected average kinetic energy of Cu$^{1+}$ ions reported in literature. The combined effects of ion loss due to the acceptance of the ion source geometry and inefficient bending are thought to result in the observed low ion transport efficiency. However, it shall be noted that the focus of this work has been to demonstrate the feasibility of a LAS with a large ablation range of up to 50 mm. A bender with smaller inter-electrode spacing and a biased aperture will improve the ion transmission efficiency, but at the cost of a reduced scanning range.\newline

\section{Conclusions}
The use of high-precision motorized kinematic mirror mounts in laser ablation is a simple and cost-effective solution for positioning the laser beam on a target, without the need to move the target inside a vacuum chamber. A 2D scan of the ion current may be used to establish a coordinate system on the surface of the target, by a simple transformation of the mirror mount motor positions. Once a coordinate system is established, different target materials may be selected and the ablated ions may be guided to another device such as a mass spectrometer. Multi-element targets can be designed from 1~mm diameter stock materials and selectively ablated for mass spectrometer calibration. The establishment of a coordinate system also allows for the monitoring and management of damage accumulation on the target surface. \newline

The PI and Thorlabs mirror mounts are capable of laser spot displacements less than $4$~$\mu$m on the target surface. However, the spatial resolution of the presented LAS system is limited by the focused laser beam diameter on the target, which is one of the many factors affecting the diameter of ablated craters. In principle, the diameter of an ablated crater is also dependent on the material and sensitive to the total optical energy deposited on the target. Systems may be designed to leverage this for a specific application. By measuring the transitions between metals, the ablation spot diameter was shown to be as low as 40-50~$\mu$m for this particular device. The calculated laser beam diameter on the target was reduced by $\approx 40\%$ by expanding the beam from 3.9(1)~mm to 9.6(1)~mm before focusing. The spatial resolution and resolving power of the LAS could be further improved by decreasing the focal length of the focusing lens.\newline

The ion transport efficiency of the ion-source assembly and quadrupole bender was measured as 2.2(7)\% for Cu, with a slight gradient in the observed in the $x$ direction. This deviation from homogeneity is likely due to a difference in proximity between the ion trajectories and the quadrupole electrodes. 

\section*{Acknowledgements}
The authors acknowledge support from the Canada First Research Excellence Fund, the Arthur B. McDonald Institute, the Canada Foundation for Innovation, the Natural Sciences and Engineering Research Council of Canada, MITACS, and the McGill Summer Undergraduate Research A\-ward program. The authors would like to thank the nEXO group at Carleton University for providing the laser used in this work.



\bibliographystyle{unsrt}

\bibliography{LAS_main.bib}

\end{document}